\documentclass{article}

\title{Isometric Entanglement of Particle Positions in Quantum Bound Systems}
\author{Robert Ducharme}

\begin{document}
\maketitle

\centerline{151 Fairhills Dr., Ypsilanti, MI 48197}
\centerline{E-mail: ducharme01@comcast.net}

\begin{abstract}
It is shown the role of a scalar potential in the Schr\"{o}dinger equation for a steady-state two-particle system is equivalent to an isometric entanglement of the position coordinates of the particles in space and time. The entangled coordinates of each particle are complex quantities related through the entangling transformation to the real positions of both particles. The transformation takes into account all of the states in the Hilbert space of the composite system. Transforming the Schr\"{o}dinger equation into these entangled coordinates eliminates the scalar potential. 
\end{abstract}

\section{Introduction}
The positions of quantum particles are correlated through both interactions and entanglement \cite{BH}. Interactions are known to lead to entanglement but entanglement can persist even in the absence of interactions. A system of two particles is entangled if the states of the system cannot be expressed as the product of the states of the individual particles \cite{ENT1}. The purpose of this paper is to show that the result of including a scalar potential in the Schr\"{o}dinger equation on bound particles is equivalent to an isometric conformal transformation \cite{RJD1,RJD2,RJD3} that entangles the position coordinates of the particles in space and time. The transformation takes account of all of the states in the Hilbert space of the bound system.

Conformal mapping \cite{ZN} is a coordinate transformation technique that has found numerous practical applications in science and engineering. Most of these applications are for two-dimensional problems but the scope of the technique is not limited to two-dimensions. Liouville's theorem \cite{DEB} in fact shows that higher dimensional conformal maps are possible but must be composed of translations, similarities, orthogonal transformations and inversions. The conformal transformation to be applied in this paper is isometric. This means the transformation used to entangle the positions of two particles in space and time does not affect the separation between the particles in either space or time.
 
Consider an isolated quantum system consisting of two bound particles each of index $k (=1,2)$ with a spatial position $x_i^k$ $(i=1,2,3)$ at time $t$. The isometric conformal transformation to be applied to this problem gives the complex coordinates for these particles to be $z_i^{k}=x_i^{k}, s=t+\imath \mathcal{\tau}(|\Psi^\prime|)$ where $\mathcal{\tau}$ is a real function of any eigensolution $\Psi^\prime$ of the quantum mechanical equation to be transformed and $\imath=\sqrt{-1}$. It is clear from inspection that this passive transformation does no more than translate both particles an equal imaginary displacement $\imath \mathcal{\tau}$ in time. The result of this translation is to entangle the position coordinates of the particles since the complex time $s$ of each particle depends on the real space and time coordinates of both. 

The aforementioned isometric conformal transformation will be introduced in section 2 of this paper. It follows from the isometric character of this transformation that the relative position of the particles $z_i (=x_i^1-x_i^2)$ is real. The complex conjugate form of $z_i$ is still written $z_i^*$ to indicate that $(z_i^*,s^*)$ and $(z_i,s)$ belong to two distinct two-particle spaces. This distinction is shown to be most evident in the computation of the partial derivatives $\partial / \partial z_i$ and $\partial / \partial z_i^*$ from the chain rule of partial differentiation as these evaluate differently in their conjugate spaces. The Cauchy-Riemann equations necessary to determining if a function transformed into the $(z_i,s)$-space has well defined partial derivatives are also given in section 2. 

The two-particle Schr\"{o}dinger equation for a quantum bound system is presented in section 3 in both real $(x_i,t)$ and complex $(z_i,s)$ coordinate systems. On comparison, it is shown that these two results are similar except for the absence of a potential term in the interaction Hamiltonian for the system in the complex coordinate system. One further topic of discussion in section 3 is that the solution $\Psi$ of the Schr\"{o}dinger equation is independent of the choice of $\Psi^\prime$ in the isometric conformal transformation. For this reason, it is often convenient to set $\Psi^\prime$ to the ground state eigensolution since this is usually the simplest option. 

The quantum harmonic oscillator is transformed into $(z_i,s)$-coordinates in section 4. It is helpful that partial derivatives with respect to spatial coordinates in the entangled space can be scaled to serve as ladder operators for raising and lowering the quantum state of the oscillator. 

\section{Entangled Position Coordinates}
The task ahead is to present an isometric conformal transformation relating two rectangular coordinate systems. This mapping is intended for application to a non-relativistic quantum bound system consisting of two particles of total energy $E$. It is convenient to express it in the form
\begin{equation} \label{eq: ict_world}
z_{i}^k = x_{i}^k, \quad s = t + \imath \frac{\hbar}{E} \ln|\Psi^\prime(x_{i}^k,t)|
\end{equation}
where $\hbar$ is Planck's constant divided by $2\pi$ and $\Psi^\prime$ is any eigensolution of the Schr\"{o}dinger equation describing the system.

It is further convenient to define center-of-mass $X_i$ and relative $x_i$ positions for the system as
\begin{equation} \label{eq: Xdef}
X_i = \frac{m_1 x_i^1 + m_2 x_i^2}{m_1+m_2}, \quad x_i=x_i^1-x_i^2 
\end{equation}
where the two particles share a common world time t. Similarly in the complex coordinate system the center-of-mass $Z_i$ and relative $z_i$ positions are
\begin{equation} \label{eq: Zdef}
Z_i = \frac{m_1 z_i^1 + m_2 z_i^2}{m_1+m_2}, \quad z_i=z_i^1-z_i^2 
\end{equation}
where the two particles share a common complex world time $s$. The key point is that in the complex coordinate system the particles have access to information about each other through $s$ that is not available in the real coordinate system through $t$. Thus, the complex coordinates may also be referred to as entangled coordinates since the idea that the separated particles have instantaneous access to information about each other is built in to the coordinate system.

For simplicity, it will be assumed from here onwards that the quantum bound system is free from external potentials so that $\Psi^\prime$ can be written in the form
\begin{equation} \label{eq: psi_form}
\Psi^\prime = \Psi_I^\prime(x_i)\exp[(\imath / \hbar)(P_i^\prime X_i-E^\prime t)] 
\end{equation}
where $\Psi_I^\prime$ has been introduced to represent the internal state of the system. Eq. (\ref{eq: ict_world}) is now readily transformed into center-of-mass and relative coordinates to give
\begin{equation} \label{eq: ict1}
z_{i} = x_{i}, \quad Z_{i} = X_{i}, \quad s = t + \imath \frac{\hbar}{E} \ln|\Psi_I^\prime(x_{i})|
\end{equation}
having assumed $|\Psi^\prime(x_i,X_i,t)|=|\Psi_I^\prime(x_{i})|$.

In the application of complex coordinates to express physical problems, there is generally going to be both a complex and a complex conjugate coordinate representation for each individual problem. In the present case, the complex conjugate of eq. (\ref{eq: ict1}) is
\begin{equation} \label{eq: ict2}
z_{i}^* = x_{i}, \quad Z_{i}^* = X_{i}, \quad s^* = t - \imath \frac{\hbar}{E} \ln|\Psi_I^\prime(x_{i})|
\end{equation}
Naturally, there must also be inverse transformations mapping the complex and complex conjugate representations of the problem back into a single coordinate system. The inverses of the transformations (\ref{eq: ict1}) and (\ref{eq: ict2}) are readily shown to be
\begin{equation} \label{eq: inv_ict1}
x_{i} = z_{i}, \quad X_{i} = Z_{i}, \quad t = s - \imath \frac{\hbar}{E} \ln|\Psi_I^\prime(z_{i})|
\end{equation}
\begin{equation} \label{eq: inv_ict2}
x_{i} = z_{i}^*, \quad X_{i} = Z_{i}^*, \quad t = s^* + \imath \frac{\hbar}{E} \ln|\Psi_I^\prime(z_{i}^*)|
\end{equation}
respectively.  

It is now interesting to investigate properties of derivatives with respect to the complex space and time coordinates. In particular, the chain rule of partial differentiation gives
\begin{equation}  \label{eq: d_dZ}
\frac{\partial}{\partial Z_i}  
= \frac{\partial X_i}{\partial Z_i} \frac{\partial}{\partial X_i}
+ \frac{\partial t}{\partial Z_i} \frac{\partial}{\partial t}
+ \frac{\partial x_{i}}{\partial Z_i} \frac{\partial}{\partial x_{i}}
= \frac{\partial}{\partial X_i} 
\end{equation}
\begin{equation} \label{eq: d_dZ_conj}
\frac{\partial}{\partial Z_i^*}  
= \frac{\partial X_i}{\partial Z_i^*} \frac{\partial}{\partial X_i}
+ \frac{\partial t}{\partial Z_i^*} \frac{\partial}{\partial t}
+ \frac{\partial x_{i}}{\partial Z_i^*} \frac{\partial}{\partial x_{i}}
= \frac{\partial}{\partial X_i} 
\end{equation}
\begin{equation}  \label{eq: d_ds}
\frac{\partial}{\partial s}  
= \frac{\partial X_i}{\partial s} \frac{\partial}{\partial X_i}
+ \frac{\partial t}{\partial s} \frac{\partial}{\partial t}
+ \frac{\partial x_{i}}{\partial s} \frac{\partial}{\partial x_{i}}
= \frac{\partial}{\partial t} 
\end{equation}
\begin{equation} \label{eq: d_ds_conj}
\frac{\partial}{\partial s^*}  
= \frac{\partial X_i}{\partial s^*} \frac{\partial}{\partial X_i}
+ \frac{\partial t}{\partial s^*} \frac{\partial}{\partial t}
+ \frac{\partial x_{i}}{\partial s^*} \frac{\partial}{\partial x_{i}}
= \frac{\partial}{\partial t} 
\end{equation}
\begin{equation} \label{eq: d_dz}
\frac{\partial}{\partial z_i} 
= \frac{\partial X_i}{\partial z_i} \frac{\partial}{\partial X_i}
+ \frac{\partial x_i}{\partial z_i} \frac{\partial}{\partial x_i}
+ \frac{\partial t}{\partial z_i} \frac{\partial}{\partial t}
= \frac{\partial}{\partial x_i} - \imath \frac{1}{\Psi_I^\prime} \frac{\partial \Psi_I^\prime}{\partial x_i} \frac{\hbar}{E} \frac{\partial}{\partial t}
\end{equation}
\begin{equation} \label{eq: d_dz_conj}
\frac{\partial}{\partial z_i^*} 
= \frac{\partial X_i}{\partial z_i^*} \frac{\partial}{\partial X_i}
+ \frac{\partial x_i}{\partial z_i^*} \frac{\partial}{\partial x_i}
+ \frac{\partial t}{\partial z_i^*} \frac{\partial}{\partial t}
= \frac{\partial}{\partial x_i} + \imath \frac{1}{\Psi_I^\prime} \frac{\partial \Psi_I^\prime}{\partial x_i} \frac{\hbar}{E} \frac{\partial}{\partial t}
\end{equation}
Note, eqs. (\ref{eq: d_dZ}), (\ref{eq: d_ds}) and (\ref{eq: d_dz}) have been obtained using eq. (\ref{eq: inv_ict1});  eqs. (\ref{eq: d_dZ_conj}), (\ref{eq: d_ds_conj}) and (\ref{eq: d_dz_conj}) are based on eq. (\ref{eq: inv_ict2}). It has also been assumed in deriving eqs. (\ref{eq: d_ds}) through (\ref{eq: d_dz_conj}) that the coordinates $(Z_i, z_i, s)$ are independent of each other as are the complex conjugate coordinates $(Z_i^*, z_i^*, s^*)$. This assumption is readily validated using eqs. (\ref{eq: d_dZ}) through (\ref{eq: d_dz_conj}) to directly evaluate derivatives of each coordinate with respect to the others. 

In further consideration of eqs. (\ref{eq: ict1}), it is convenient to write $s=t+i\tau$ where 
\begin{equation}
\tau = \frac{\hbar}{E} \ln|\Psi_I^\prime(x_{i})|
\end{equation}
The requirement for a continuously differentiable function $f(s)=g(t,\tau)+ih(t,\tau)$ to be holomorphic is then for the real functions g and h to satisfy the set of Cauchy-Riemann equations
\begin{equation}
\frac{\partial g}{\partial \tau}=\frac{\partial h}{\partial t} \quad
\frac{\partial h}{\partial \tau}=-\frac{\partial g}{\partial t} 
\end{equation}
or equivalently
\begin{equation} \label{eq: creq}
\frac{\partial^2 f}{\partial t^2} + \frac{\partial^2 f}{\partial \tau^2} = 0
\end{equation}
It is thus concluded that a function $\Psi(x_i,t)$ will also have an equivalent holomorphic form $\theta(z_i)f(s)$ in the complex $(z_\mu, s)$-coordinate system providing it is separable and $f$ satisfies eq. (\ref{eq: creq}). Here, it is understood that the domain of the Cauchy-Riemann equations in this problem is the complex plane containing s. The Cauchy-Riemann equations put no restriction at all on the form of the function $\theta(z_i)$ since $z_i$ and $s$ are independent coordinates and $z_i$ belongs to a real three-dimensional space.

\section{Quantum Bound Systems}
The Schr\"{o}dinger equation determining the wavefunction $\Psi$ for two particles bound through a mutual scalar potential $V(x_i)$ is
\begin{equation}\label{eq: Schrod_2part}
\frac{\hbar^2}{2m_c} \frac{\partial^2 \Psi}{\partial X_i^2} + \frac{\hbar^2}{2m_r}\frac{\partial^2 \Psi}{\partial x_i^2} + (E-V) \Psi = 0
\end{equation}
(see ref. \cite{DFL}) with
\begin{equation}\label{eq: op_time}
\imath \hbar \frac{\partial \Psi}{\partial t} = E\Psi.
\end{equation}
where $m_c=m_1+m_2$ and $m_r=m_1m_2/m_c$. Assuming the product solution $\Psi=\Psi_E(X_i)\Psi_I(x_i)\exp(-\imath Et / \hbar)$ this expression can be separated into the two equations
\begin{equation}\label{eq: Schrod_com}
\frac{\hbar^2}{2m_c} \frac{\partial^2 \Psi}{\partial X_i^2} + (E-E_n) \Psi = 0
\end{equation}
\begin{equation}\label{eq: Schrod_rel}
\frac{\hbar^2}{2m_r}\frac{\partial^2 \Psi}{\partial x_i^2} + (E_n-V) \Psi = 0
\end{equation}
where $E_n$ is the total internal energy of the bound system.

In seeking to transform eq. (\ref{eq: Schrod_rel}) into entangled coordinates (\ref{eq: ict1}), the first step is to note that $\Psi^\prime$ must be an eigensolution of the equation
\begin{equation}\label{eq: Schrod_sub}
\frac{\hbar^2}{2m_r}\frac{\partial^2 \Psi^\prime}{\partial x_i^2} + (E_m-V) \Psi^\prime = 0
\end{equation}
identical in form to eq. (\ref{eq: Schrod_rel}). Eqs. (\ref{eq: d_dz}) and (\ref{eq: d_dz_conj}) are readily combined to give the expression
\begin{equation} \label{eq: d2_dz2}
\frac{\partial^2\Psi}{\partial z_i^*\partial z_i} = \frac{\partial^2 \Psi}{\partial x_i^2} - \frac{\Psi}{\Psi^\prime} \frac{\partial^2 \Psi^\prime}{\partial x_i^2}
\end{equation}
having made use of eq. (\ref{eq: op_time}). Inserting eqs. (\ref{eq: Schrod_rel}) and (\ref{eq: Schrod_sub}) into eq. (\ref{eq: d2_dz2}) leads to
\begin{equation}\label{eq: Schrod_rel_entang}
\frac{\hbar^2}{2m_r}\frac{\partial^2 \Psi}{\partial z_i^*\partial z_i} + (E_n-E_m) \Psi = 0
\end{equation}
The transformation of the Schr\"{o}dinger equation (\ref{eq: Schrod_2part}) into entangled coordinates may be completed using eqs. (\ref{eq: d_dZ}), (\ref{eq: d_dZ_conj}) and (\ref{eq: Schrod_com}) to yield
\begin{equation}\label{eq: Schrod_com_entang}
\frac{\hbar^2}{2m_c} \frac{\partial^2 \Psi}{\partial Z_i^* \partial Z_i} + (E-E_n) \Psi = 0
\end{equation}
describing the motion of the bound system in center-of-mass coordinates. Eqs. (\ref{eq: d_ds}) and (\ref{eq: op_time}) also give the related expression
\begin{equation} 
\imath \hbar \frac{\partial \Psi}{\partial s} = E\Psi.
\end{equation}
for the total energy of the bound system.

It is significant that the potential energy term $V(x_i)$ in eq. (\ref{eq: Schrod_rel}) has no counterpart in eq. (\ref{eq: Schrod_rel_entang}). In moving from real to complex coordinates, the concept of a position-dependent potential therefore vanishes from the interaction Hamiltonian for any bound system. The explanation for the particles remaining in the bound state in the absence of an interaction potential is that their positions are entangled. Specifically, the particles evolve in a complex time coordinate $s$ such that two particles sharing a common complex time $s$ have correlated spatial positions in space as well as a common real time $t$. 

\section{The Harmonic Oscillator}
The harmonic oscillator potential is
\begin{equation} \label{eq: potential_ho} 
V=\frac{1}{2}m_r \omega^2 r^2
\end{equation} 
where $\omega$ is the spring constant of the oscillator and $r=|x_i|$. Inserting this into eq. (\ref{eq: Schrod_rel}) gives
\begin{equation} \label{eq: psi1} 
\Psi(x_i,t) = \frac{(m_r \omega / \pi \hbar)^{3/4}}{\sqrt{2^n l_1!l_2!l_3!}}  H_{l_1}(\xi_1)H_{l_2}(\xi_2)H_{l_3}(\xi_3)\exp\left( -\frac{\imath 2Et + m_r\omega r^2}{2\hbar} \right)
\end{equation}
where $\xi_i=\sqrt{\frac{m_r \omega}{\hbar}}x_i$, $H_{l_j}$ are Hermite polynomials and $l_1,l_2,l_3$ are positive integers. The corresponding energy eigenvalues are
\begin{equation} \label{eq: energy} 
E_n = \hbar \omega \left(\frac{3}{2} + n \right)
\end{equation}
where $n=l_1+l_2+l_3$. 

It is convenient to choose $\Psi_I^\prime$ in eq. (\ref{eq: ict1}) to be the ground state solution 
\begin{equation} \label{eq: psi_gs} 
\Psi_I^\prime = \left(\frac{m_r \omega}{\pi \hbar}\right)^{3/4}\exp \left( -\frac{m_r \omega r^2}{2\hbar} \right)
\end{equation}
giving the complex world time for the harmonic oscillator to be
\begin{equation}
s = t + \imath\frac{\hbar}{E}\left[ \frac{3}{4}\ln\left( \frac{m_r \omega}{\pi \hbar}\right) - \frac{m_r\omega r^2}{2\hbar} \right]
\end{equation}
Further, putting eq. (\ref{eq: psi_gs}) into eqs. (\ref{eq: d_dz}) and (\ref{eq: d_dz_conj}) gives
\begin{equation} \label{eq: complexDiff6}
\frac{\partial}{\partial z_i} 
= \frac{\partial}{\partial x_i} + \frac{m_r \omega}{\hbar} x_i 
\end{equation}
\begin{equation} \label{eq: complexDiff7}
\frac{\partial}{\partial z_i^*} 
= \frac{\partial}{\partial x_i} - \frac{m_r \omega}{\hbar} x_i
\end{equation}
showing the ladder operators for lowering and raising the state of the oscillator can be expressed in the compact form
\begin{equation}  \label{ladder_nr}
\hat{a}_i = \sqrt{\frac{\hbar}{2m_r\omega}}\frac{\partial}{\partial z_i}, \quad \hat{a}_i^\dag = -\sqrt{\frac{\hbar}{2m_r\omega}}\frac{\partial}{\partial z_i^*}
\end{equation} 
Now the internal energy corresponding to the eigenstate (\ref{eq: psi_gs}) is
\begin{equation} \label{eq: energy_gs} 
E_0 = \frac{3}{2} \hbar \omega 
\end{equation}
Thus, inserting eqs. (\ref{ladder_nr}) and (\ref{eq: energy_gs}) into eq. (\ref{eq: Schrod_rel_entang}) gives
\begin{equation} \label{eq: ladop_eq} 
\left(\hat{a}_i^\dag \hat{a}_i+\frac{3}{2} \right)\hbar \omega \Psi = E_n \Psi
\end{equation} 
This is the familiar ladder operator form of the Schr\"{o}dinger equation for the harmonic oscillator.

The oscillator function (\ref{eq: psi1}) is readily transformed into  $(z_i,s)$-coordinates using eqs. (\ref{eq: ict1}) to give
\begin{equation} \label{eq: psi_entang} 
\Psi(z_i,s) = \frac{(m_r \omega / \pi \hbar)^{3/4}}{\sqrt{2^n l_1!l_2!l_3!}}  H_{l_1}(\zeta_1)H_{l_2}(\zeta_2)H_{l_3}(\zeta_3)\exp\left( -\frac{\imath Es }{\hbar} \right)
\end{equation}
and $\zeta_i = \sqrt{\frac{m_r \omega}{\hbar}}z_i$. It is notable that eq. (\ref{eq: psi1}) and (\ref{eq: psi_entang}) are similar except that eq. (\ref{eq: psi_entang}) does not contain a gaussian term. It is also notable that eq. (\ref{eq: psi_entang}) is a continuously differentiable solution of the Cauchy-Riemann equations (\ref{eq: creq}) thus demonstrating that the oscillator function $\Psi(z_i, s)$ is holomorphic.

\section{Conclusion}
It is concluded that there are two equivalent models of how the positions of a pair of particles are correlated in a non-relativistic quantum bound system. These are the interaction and entanglement models. In the interaction model, the particles continuously influence each other through potential fields that are assumed to surround each of them. By contrast, in the entanglement model there are no potential fields. Instead the time of each particle is treated as a complex quantity that depends on the real positions of both the particles in space and time. Particles sharing a common complex time also have correlated positions. The interaction and entanglement models are related through an isometric conformal transformation.

\end{document}